\documentclass[pre]{revtex4}
\usepackage{graphicx}
\usepackage{latexsym}
\usepackage{amsmath}
\usepackage{amssymb}
\usepackage{amsfonts}
\usepackage{amsthm}

\begin{document}
\title{A new (2+1) dimensional integrable evolution 
equation for an ion acoustic wave in a magnetized plasma}
\author{ Abhik Mukherjee
\footnote{abhik.mukherjee@saha.ac.in}
, M.S. Janaki
\footnote{ms.janaki@saha.ac.in}
, Anjan Kundu
\footnote{anjan.kundu@saha.ac.in}
}
\affiliation{Saha Institute of Nuclear Physics\\
 Kolkata, INDIA}

\begin{abstract} 
A new, completely integrable, two dimensional evolution equation is derived for  an ion acoustic wave
propagating in a magnetized, collisionless plasma. 
The equation is a multidimensional generalization of a modulated wavepacket with weak transverse propagation which has resemblance to nonlinear Schrodinger equation and has   a connection to  Kadomtsev- Petviashvili (KP) equation through a constraint relation. Higher soliton solutions of the equation are derived through Hirota bilinearization procedure and an exact lump solution is calculated exhibiting 2D structure. Some mathematical properties  demonstrating the completely integrable nature of this equation are described.   Modulational instability using nonlinear frequency correction is  derived and the corresponding growth rate is calculated which shows the directional asymmetry 
of the system. The discovery of this novel (2+1) dimensional integrable NLS type equation for a magnetized plasma should pave a new direction of research in the field.

\end{abstract}
\maketitle

\smallskip


\section{Introduction}	Active research on  nonlinear phenomena in plasma physics 
	has grown extensively and gained much importance over the past 
	few decades due to failure of linear theory in 
	explaining phenomena related to large  amplitude waves, 			
wave- particle, wave-wave interactions etc \cite{Chen}.
	However, the complexity of the associated nonlinear partial differential equations makes the system 
 less well understood in most of the cases. There lies the importance of the integrable nonlinear equations
 because of their rich analytical
beauty and availability of generalized mathematical techniques for solving them \cite{Lakshmanan}.
Washimi and Taniuti \cite{Washimi} were first to derive the completely integrable Korteweg de Vries (KdV) equation
for small but finite amplitude ion acoustic solitary waves for collision less plasma composed of cold ions and hot electrons. Since then the plasma physics community has been actively involved
in  nonlinear phenomena related structures such as solitons, shocks, instabilities, wave-wave and wave-particle
interactions etc. The first experimental observation of ion acoustic soliton has been made by
Ikezi et.al \cite{Ikezi,Tran}. Since then the KdV model has been used extensively in various branches like
dusty plasma\cite{Cousens,Mamun,Bacha}, Bose Einstein gravitationally condensed gas\cite{NC},
weakly relativistic magnetized plasma\cite{Malik}, non-thermal plasma\cite{Verheest}, dense plasma with degenerate
electron fluids\cite{Shukla} in planar as well as nonplanar geometry\cite{Huang,Maxon} and also in other branches. Other nonlinear equations like
Boussinesq equation\cite{Tran}, Benjamin-Bona-Mahony (BBM) equation\cite{Broer}, which are not integrable also find their applications  in plasma physics. 
It is well known that nonlinear wave propagation is generally subject to an amplitude modulation due
to carrier wave self interaction resulting in a slow modulation of monochromatic plane wave leading to the 
formation of an envelope soliton, which may be described by the Nonlinear Schr$\ddot{o}$dinger( NLS) equation,
 also a completely
integrable system. This equation is also investigated extensively in various areas of plasma systems
like dusty plasma\cite{Amin,Moslem},  multicomponent plasma\cite{Sabry},
in explaining rogue waves\cite{Moslem,Langmuir,hydromagnetic}, relativistic laser plasma interactions\cite{Javan}
 and  various other fields. Peregrine soliton of NLS equation which is used to describe rogue waves is 
experimentally observed in a multicomponent plasma with negative ions\cite{Bailung}. A complex Ginzberg-Landau
equation is derived in compressional dispersive Alfvenic waves in a collisional magnetoplasma\cite{hydromagnetic}
which reduces to standard NLS equation in a collisionless plasma.
The discussed equations are all (1+1) dimensional, but in practical circumstances the waves observed in
laboratory and space are certainly not bounded in one dimension. Franz\cite{Franz} et.al have shown that a 
purely 1D model cannot account for the observed features in the auroral region, especially at higher 
polar altitudes. 
The best known 2D generalization of KdV equation are Kadomtsev- Petviashvili (KP) equation and
Zakharov- Kuznetsov(ZK) equation. A completely integrable generalization of the KdV equation is the KP equation
\cite{KP} which has also been used in various branches of plasma such as  inhomogeneous plasma with finite
temperature drifting ions\cite{Dahiya}, ultracold quantum magnetospheric plasma\cite{Mushtaq}, electron
positron ion plasma \cite{Shahmansouri} and also in other areas. The  stability of their solutions 
under transverse perturbations was also studied \cite{Kako,Mushtaq}. The ZK equation \cite{ZK}
which is  more isotropic in transverse direction was first derived for describing weakly nonlinear ion
acoustic waves in strongly magnetized lossless plasma in 2D\cite{Zakharov2}. It was also reported that this equation 
is not integrable under inverse scattering method\cite{Bhimsen,Infeld} and till date only three 
polynomial conservation laws have been given \cite{Infeld2, Bhimsen2}. This equation was also explored vastly
in the last few decades \cite{Zhen,Saini,Labany,Bains} and higher dimensional solitons were derived
\cite{Mace, Wazwaz}. A 2D generalization of NLS equation is DS equation which was also derived
for electrostatic ion waves \cite{Nishinari}, electron acoustic wave \cite{Ghosh}, space and laboratory 
dusty plasma \cite{Annou}, and in cylindrical geometry \cite{Xue}. For special choice of coefficients
DS equation converges to DS1 equation which is analytically integrable and admits dromion solutions 
with localized structure in higher dimensions \cite{Ghosh, Nishinari, Annou, Numanal, Boiti}.
But in case of DS1 equation additional fields are coupled in the interacting term which could be related to
the basic fields only through nonlocal transformations.
Hence due to the presence of a few integrable equations (both in 1D and 2D) in plasma systems, there is always
a requirement for the discovery of new integrable equations (specially in 2D). 

In this work we have derived
a completely integrable 2D nonlinear evolution equation in lossless magnetized plasma
 with asymmetric scaling on transverse variable. This equation
involving only local interactions of dependent variables
 was derived earlier in 
hydrodynamic system \cite{MyRogue} . 
 The 2D generalizations of NLS equation available in the literature involve either
nonlocal interactions, or are
 non-integrable, whereas
 the equation presented here
is the completely integrable, local, (2+1) dimensional generalization of NLS 
equation which however possesses many properties similar to (1+1) dimensional NLS equation.

The paper is organized as follows. The derivation of the (2+1) dimensional, integrable, evolution equation 
for an ion acoustic wave in a magnetized plasma is given in section II
with asymmetric scaling on transverse variables.
Nonlinear frequency correction and modulation instability of the evolution equation
are discussed in Section III indicating strong directional preference
and asymmetry in transverse direction.
 The connection of the equation with KP equation is discussed in section IV.
 One and two soliton solutions of this equation using Hirota bilinearization method are given in section V and
 in section VI, an exact static lump solution carrying two free parameters is presented.
 Other integrable properties of this novel system  like Lax pair,
infinite conserved quantities etc. are included in Section VII.
  Conclusive remarks and Appendix are
given in Section VIII and Section IX respectively, followed by the bibliography.

\section{
Derivation of two dimensional integrable equation for electrostatic waves propagating in a magnetized plasma}

A new  two dimensional  integrable evolution equation for the propagation of nonlinear waves
in magnetized plasma is derived in this section. We  consider the 
propagation of electrostatic waves in a magnetized plasma with the magnetic
 field $B= B_0 \hat{e_z}$, where $B_0$ is a constant.  For situations where plasma
 pressure is much smaller than the magnetic pressure, plasma wave excitation is
 in general electrostatic. The focus is on a plasma  composed of two components, ions 
and electrons that are described by fluid equations under collision-free conditions in 
the cartesian coordinates, which include conservation of mass and momentum together with Poisson's equation given by

\begin{equation}
\frac{\partial n}{\partial t}+\overrightarrow{\bigtriangledown}\cdot(n \overrightarrow v ) =0
,\ \
\frac{\partial \overrightarrow v}{\partial t}+(\overrightarrow v\cdot\overrightarrow{\bigtriangledown})\overrightarrow v +
\overrightarrow \bigtriangledown \phi - \alpha(\overrightarrow v \times \overrightarrow b) =0, \ \
\bigtriangledown^2 \phi =  n_e - n, 
n_e = \exp(\phi) \label{basic}
\end{equation} 

 where $n_e, n, v, B,\phi$ are electron, ion densities and ion fluid velocity,
 magnetic field and electrostatic potential respectively and $\alpha $ is a dimensionless
 parameter given by  $ \omega_{ci}/\omega_{pi}$. For convenience, we have used the following normalization resulting in dimensionless parameters:   electron  and ion densities  normalized by $n_0$, coordinates
 by electron Debye length $\lambda_{De} = v_{te}/\omega_{pe}$, fluid velocity by the acoustic speed $c_s = \sqrt{k_BT_e/m_i}$; time  by ion plasma period $\omega_{pi}^{-1}$, and magnetic field $B$ by  $B_0=m_e\omega_{pi}/e$, where $v_{te}, \omega_{pe},\omega_{pi},\omega_{ci}$ are
the electron thermal velocity, electron and ion plasma frequencies and ion cyclotron frequency
respectively.

In the above equations, the ions are assumed to be cold and on the slow ion time scale, the
 electrons are assumed to be in local thermodynamic equilibrum.
When the electron inertia is neglected, the electrons can be considered to follow a Boltzmann distribution if  the propagation vector has a small component along the magnetic field, such that
the angle $\chi$ between the wave vector normal to the magnetic field and the 
wave vector is larger than $\sqrt{m_e T_i/m_i T_e}$, so that as a special case we can take 
 $k_z \rightarrow 0 $.  This enables us to consider propagation perpendicular to the magnetic field with the
 wave vector $k = (k_x, k_y,0)$. 
 The linear  propagation of electrostatic ion cyclotron waves 
 propagating perpendicular to the magnetic field is governed by the dispersion relation
$$ \omega^2 = k^2c_s^2/(1+k^2\lambda_{De}^2) + \omega_{ci}^2 $$
where $k^2=k_x^2+k_y^2$.

In order to derive the nonlinear evolution equation  governing the propagation of the 
electrostatic ion cyclotron waves, we  assume perturbation of the form $\sim$ 
exp[i($\overrightarrow{k}$ $\cdotp$ $\overrightarrow {r}$ - $\omega$ t)],  and  adopt  the
reductive 
perturbation expansion technique.  All the physical quantities
are expanded about their equilibrium values  as-
\begin{equation}
n= 1 + \sum_{m=1}^{\infty}\epsilon^m \sum_{l=-m}^{m}n_{l}^{(m)} exp[i l(\overrightarrow{k} \cdotp \overrightarrow {r}-\omega t)]
\label{n}\end{equation}
\begin{equation}
\phi =  \sum_{m=1}^{\infty}\epsilon^m \sum_{l=-m}^{m}\phi_{l}^{(m)}  exp[i l(\overrightarrow{k} \cdotp \overrightarrow {r}-\omega t)]
\label{phi}\end{equation}
\begin{equation}
v_{j} =  \sum_{m=1}^{\infty}\epsilon^m \sum_{l=-m}^{m} v_{j l}^{(m)}  exp[i l(\overrightarrow{k} \cdotp \overrightarrow {r}-\omega t)]
\label{v}\end{equation}
where $j$  denotes the $x$ and $y$ components of ion velocities.
We  have introduced the following stretched variables with asymmetric scaling on
transverse direction as
 \begin{eqnarray}
\xi= \epsilon (x- M_{x} t), \ \
\eta= \epsilon ^2 y, \ \
\tau= \epsilon^3 t
\label{scaling}\end{eqnarray} 
where $M_{x}$ is the group velocity in the $x$ axis.  The scaling used here is different from the scaling
involved in the derivation of Davey-Stewartson equation that has a symmetric dependence on  all the space variables. The  stretching  in this case is asymmetric with respect to one of the space variables.
Such a situation may arise in some experimental scenario where
 there is a  possibility of weak dependence in one of the directions.

Transforming all independent variables by equation (\ref{scaling}), we
expand equations (\ref{basic}) and carry out a systematic
balancing of terms at each order of $\epsilon$.
The coefficients appearing at different orders are all given at the appendix.

At $\bf{\epsilon: l=1}$ order we get
\begin{equation}
\phi_{1} = K_{1}n_{1}^{(1)}, \ \
v_{x1}^{(1)} =  A_{1}^{(1)} n_{1}^{(1)}, \ \
v_{y1}^{(1)} = B_{1}^{(1)} n_{1}^{(1)}
 \end{equation}

Combining the above expressions leads to the linear dispersion relation for the ion acoustic wave-
\begin{equation}
\omega^2 = |k|^2 K_{1} + \alpha^2, K_1 = 1/(1+{\mid k \mid}^2)
\end{equation}

Similiarly at  $\bf{\epsilon^2: l=0;}$ we get,

\begin{equation}
v_{x0}^{(2)} =  A_{0}^{(2)} |n_{1}|^{2},\ \
v_{y0}^{(2)} = B_{0}^{(2)} |n_{1}|^{2}, \ \
\phi_{0}^{(2)}= - K_{1}^2|n_{1}^{(1)}|^2 + n_{0}^{(2)}
\end{equation}

At  $\bf{\epsilon^2: l=1;}$ we obtain,

\begin{equation}
 \phi_{1}^{(2)}=  K_{1} n_{1}^{(2)} + 2 i k_{x}K_{1}^2 \frac{\partial n_{1}^{(1)}}{\partial \xi},\ \
v_{x1}^{(2)} =  A_{1}^{(2)} n_{1}^{(2)} + B_{1}^{(2)} \frac{\partial n_{1}^{(1)}}{\partial \xi},\ \
v_{y1}^{(2)} = C_{1}^{(2)} n_{1}^{(2)} + D_{1}^{(2)} \frac{\partial n_{1}^{(1)}}{\partial \xi}
 \end{equation}

The group velocity along x axis, $M_{x}$, can be found out from this order of calculation as 
\begin{equation}
 M_{x} = \frac{(A_{1}^{(1)} ( \omega^2- \alpha^2 ) + k_{x} K_{1} \omega - i K_{1} k_{y} \alpha - 
   2 K_{1}^2 k_{x} \omega |k|^2}{(-\alpha^2 + \omega^2 + i B_{1}^{(1)} k_{x} \alpha + 
   A_{1}^{(1)} k_{x} \omega - i A_{1}^{(1)} k_{y} \alpha + B_{1}^{(1)} k_{y}  \omega)} 
\end{equation}

At  $\bf{\epsilon^2: l=2;}$ 

\begin{equation}
 \phi_{2}^{(2)}=  D_{2}^{(2)} (n_{1}^{(1)})^2,\ \
v_{x2}^{(2)} =  A_{2}^{(2)} (n_{1}^{(1)})^2,\ \
v_{y2}^{(2)} = B_{2}^{(2)} (n_{1}^{(1)})^2,\ \
n_{2}^{(2)} = C_{2}^{(2)} (n_{1}^{(1)})^2
\end{equation}

At $\bf{\epsilon^3: l=1;}$ order, an NLS-type equation (space co-ordinate $\eta$
replacing the time co-ordinate) is obtained as-

\begin{equation}
i A_{1}^{(3)} \frac{\partial n_{1}^{(1)}}{\partial \eta} + B_{1}^{(3)} \frac{\partial^2 n_{1}^{(1)}}{\partial \xi^2}
+ C_{1}^{(3)} |n_{1}^{(1)}|^{2} n_{1}^{(1)} = 0
\label{spaceNLS}\end{equation} 

The above space-type NLS equation has resulted because in the
present work, we have scaled the transverse variable $y$ in the
 same way as time is scaled in the derivation of NLS equation. 

At $\bf{\epsilon^3: l=0;}$ order we find $n_0^{(2)} = 0$ since $n \rightarrow 0 $ as $\xi\rightarrow \infty$.
The other quantities determined are-
\begin{equation}
 v_{x0}^{(3)} = A_{0}^{(3)} n_{1}^{(1)*}\frac{\partial n_{1}^{(1)}}{\partial \xi}+
 B_{0}^{(3)} n_{1}^{(1)}\frac{\partial n_{1}^{(1)*}}{\partial \xi}+
 C_{0}^{(3)} n_{1}^{(1)*} n_{1}^{(2)} + C_{0}^{(3)} n_{1}^{(1)} n_{1}^{(2)*}
\end{equation}
\begin{equation}
 v_{y0}^{(3)} = E_{0}^{(3)} n_{1}^{(1)*}\frac{\partial n_{1}^{(1)}}{\partial \xi}+
 F_{0}^{(3)} n_{1}^{(1)}\frac{\partial n_{1}^{(1)*}}{\partial \xi}+
 G_{0}^{(3)} n_{1}^{(1)*} n_{1}^{(2)} + H_{0}^{(3)} n_{1}^{(1)} n_{1}^{(2)*}
\end{equation}
\begin{equation}
 n_{0}^{(3)} = I_{0}^{(3)} n_{1}^{(1)*}\frac{\partial n_{1}^{(1)}}{\partial \xi}+
 J_{0}^{(3)} n_{1}^{(1)}\frac{\partial n_{1}^{(1)*}}{\partial \xi}+
 K_{0}^{(3)} n_{1}^{(1)*} n_{1}^{(2)} + L_{0}^{(3)} n_{1}^{(1)} n_{1}^{(2)*}
\end{equation}
\begin{equation}
 \phi_{0}^{(3)} = M_{0}^{(3)} n_{1}^{(1)*}\frac{\partial n_{1}^{(1)}}{\partial \xi}+
 P_{0}^{(3)} n_{1}^{(1)}\frac{\partial n_{1}^{(1)*}}{\partial \xi}+
 N_{0}^{(3)} n_{1}^{(1)*} n_{1}^{(2)} + Q_{0}^{(3)} n_{1}^{(1)} n_{1}^{(2)*}
\end{equation}

Detailed mathematical forms of all the coefficients occuring in the above equations are given in the Appendix. Similarly the 
$\bf{\epsilon^3: l=2;}$ order quantities like $v_{x2}^{(3)}, v_{y2}^{(3)}$ etc can be determined by the same 
procedure but the exact expressions cannot be given in view of their
extreme cumbersome nature.

Finally at $\bf{\epsilon^4: l=1;}$  order a two dimensional evolution equation is obtained in the form

\begin{eqnarray}
 i A_{1}^{(4)} \frac{\partial n_{1}^{(1)}}{\partial \tau} +B_{1}^{(4)} \frac{\partial^2 n_{1}^{(1)}}{\partial \xi \partial \eta}
+ i C_{1}^{(4)} \frac{\partial^3 n_{1}^{(1)}}{\partial \xi^3}+
i D_{1}^{(4)}(n_{1}^{(1)})^2 \frac{\partial n_{1}^{(1)*}}{\partial \xi}
+i E_{1}^{(4)}|n_{1}^{(1)}|^2 \frac{\partial n_{1}^{(1)}}{\partial \xi}\nonumber\\
+F_{1}^{(4)} |n_{1}^{(1)}|^2 n_{1}^{(2)}+  G_{1}^{(4)}(n_{1}^{(1)})^2 n_{1}^{(2)*}
+i H_{1}^{(4)} \frac{\partial n_{1}^{(2)}}{\partial \eta}+
  I_{1}^{(4)} \frac{\partial^2 n_{1}^{(2)}}{\partial \xi^2}
 = 0 &&
\label{2dE}\end{eqnarray}

where the coefficients $A_{1}^{(4)}$ - $I_{1}^{(4)}$, which are real constants dependent on parameters $k_x, k_y$ and
$\alpha$, are too cumbersome to be expressed in an explicit form. This is a general two dimensional
non-integrable equation of two dependent variables $n_1^{(1)}$ and $n_1^{(2)}$. If it is assumed that the term 
$n_1^{(2)}$ depends on $n_1^{(1)}$ like the other terms as $v_{x1}^{(1)}, v_{y1}^{(1)} \sim n_1^{(1)}$,
$v_{x2}^{(2)}, v_{y2}^{(2)}, n_2^{(2)} \sim (n_1^{(1)})^2$ etc, then the only possible consistent relation between 
$n_1^{(1)}$ and $n_1^{(2)}$ would be 
$n_1^{(2)}\sim \frac{\partial n_1^{(1)}}{\partial \xi}$. Hence we consider $n_1^{(2)}= i P_1 \frac{\partial n_{1}^{(1)}}{\partial \xi}$
where $P_1 $ is a constant dependent on $k_x, k_y, \alpha$. Now using (\ref{spaceNLS}) in (\ref{2dE})
we see that the general nonintegrable equation (\ref{2dE}) turns into the form
\begin{equation}
 i C_0 \frac{\partial n_{1}^{(1)}}{\partial \tau} + C_1 \frac{\partial^2 n_{1}^{(1)}}{\partial \xi \partial \eta} +
2 i C_2 n_{1}^{(1)}(n_{1}^{(1)} \frac{\partial n_{1}^{(1)*}}{\partial \xi} -n_{1}^{(1)*}
\frac{\partial n_{1}^{(1)}}{\partial \xi})
= 0 \label{2dnls1}
\end{equation}
for the choice of the constant 
\begin{equation}
 P_1 = \frac{[\frac{3 C_1^{(3)} C_1^{(4)}}{B_1^{(3)}}-D_1^{(4)}- E_1^{(4)} ]}
 {[F_1^{(4)}- G_1^{(4)} - \frac{3 I_1^{(4)} C_1^{(3)}}{B_1^{(3)}}]} \label{P1}
\end{equation}
where $C_0, C_1, C_2$ depends on the parameters $k_x, k_y, \alpha$.
 In case of the multidimensional extension of modulated ion acoustic 
wave by Nishinari\cite{Nishinari},  general multidimensional coupled 
equations were obtained which were converted to the integrable DS1 equation for the specific choice 
 ($k_x \rightarrow 0$, $\alpha_1\alpha_2 > 0$ and all the variables independent of $\zeta$ \cite{Nishinari}).
 Similarly,  in our case, for the specific choice of $P_1$ as given in (\ref{P1}),
 the general nonlinear nonintegrable equation becomes
 completely integrable (\ref{2dnls1}).
 As the explicit representation of the coefficients $C_0,C_1$ and $C_2$ are too cumbersome, we will
show their  behavior graphically. In Figure (1), we plot the variation of the coefficients
$C_0, C_1$ and $C_2$ with $\alpha$ for $k_x=1, k_y=1$.
\begin{figure}[!h]
\noindent {\bf a.} 
\includegraphics[width=7.cm, angle=0]{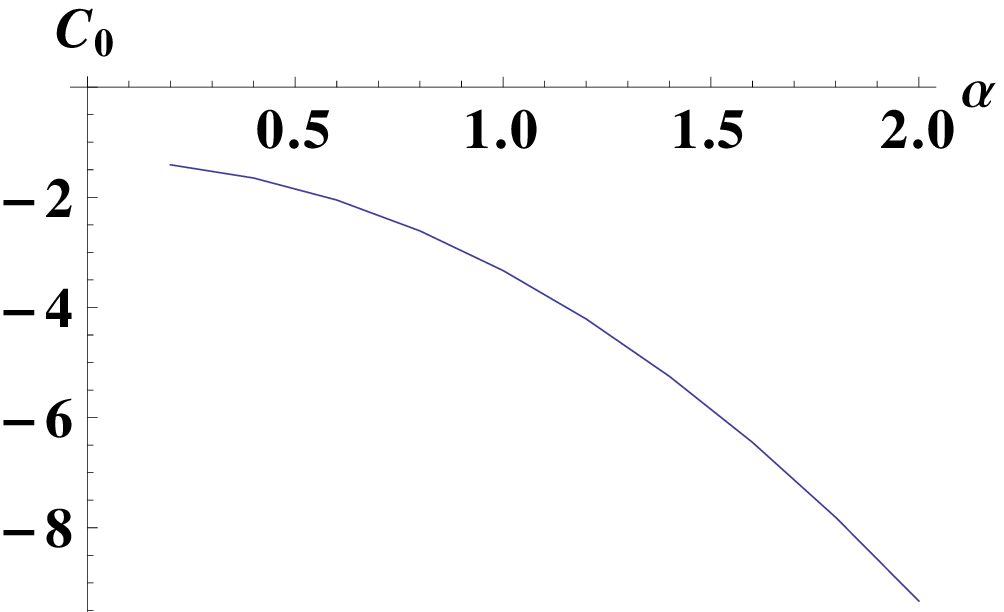}
\\
\noindent{\bf b.}
{
\includegraphics[width=7cm, angle=0]{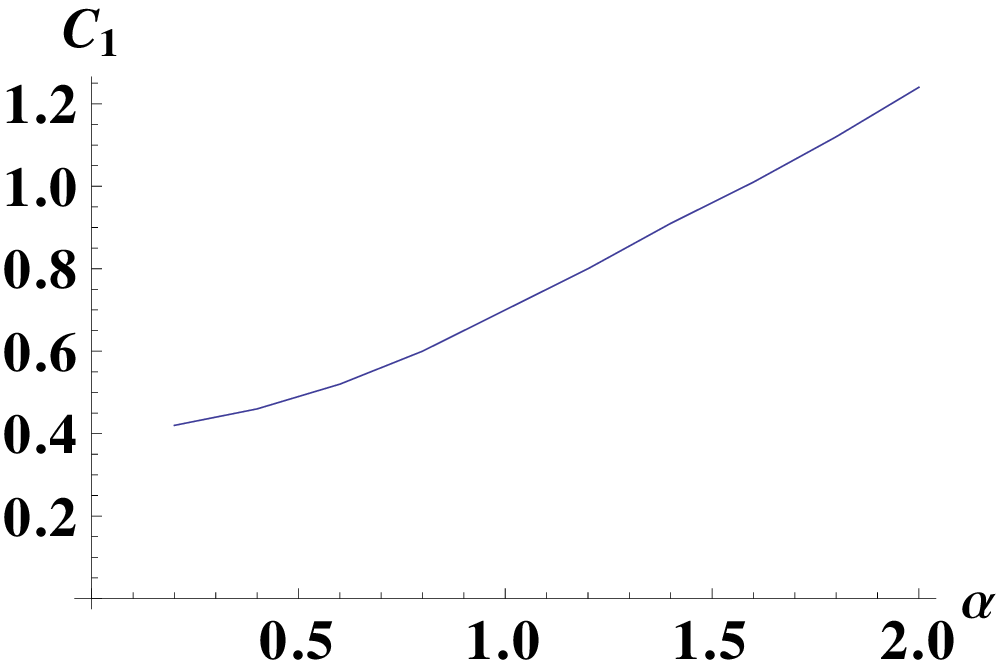}  
}
\\
\noindent {\bf c.}
 \includegraphics[width=7.cm, angle=0]{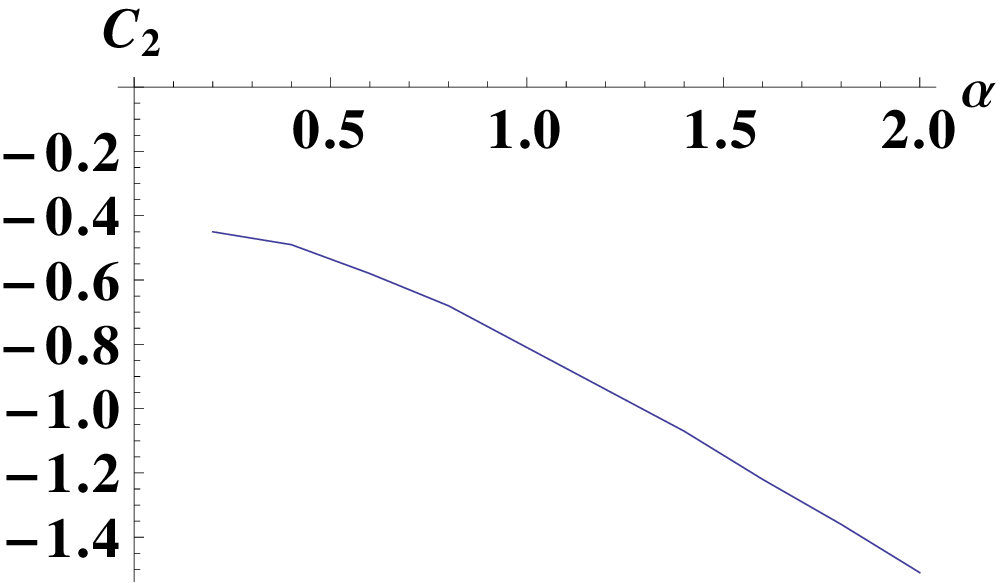}

\caption
{
Variation of coefficients $C_0,C_1,C_2$ with $\alpha$
}
\end{figure}

Now rescaling the variables $\tau, \xi$ and $\eta$ in (\ref{2dnls1}) we get 
\begin{equation}
 i \frac{\partial n_{1}^{(1)}}{\partial_{t}} +  \frac{\partial^2 n_{1}^{(1)}}{\partial_{x}\partial_{y}} +
2 i n_{1}^{(1)}(n_{1}^{(1)} \frac{\partial n_{1}^{(1)*}}{\partial_{x}} -n_{1}^{(1)*} \frac{\partial n_{1}^{(1)}}{\partial_{x}}) = 0
\label{2dsInls}\end{equation}
 and renaming $n_{1}^{(1)}$ as $u$ it gives
\begin{equation}
 iu_{t}+u_{xy}+2iu(uu_{x}^*-u^*u_{x})=0
\label{2dnls}
\end{equation}
which is our new (2+1) dimensional completely integrable evolution equation that has been obtained at a higher 
perturbation order compared to the NLS equation, hence expected to address weaker effects. Similar equation  was derived in the context of water 
waves\cite{MyRogue} in order to model oceanic rogue waves.
It has structural similarity  with the NLS equation, where the nonlinearity   comes from the ponderomotive  force  that depends on the square of modulus and not on the phase of $u$.  The present equation(\ref{2dnls}) has slightly different characteristics, with the nonlinear potential dependent on the square of modulus of the wave profile as well as on the x-derivative
of phase.  The dispersive term, is a cross derivative term dependent on both longitudinal and transverse directions.

The study of propagation of modulated ion acoustic waves in  the presence of a  magnetic field
has been extensively done using the NLS equation that restricts the study to one dimension. 
  A multidimensional generalization of the NLS equation for a modulated ion acoustic wave packet
 propagating in a magnetized plasma leads to the 
Davey Stewartson equation \cite{Nishinari,Annou}. However, since all the spatial directions 
 are scaled symmetrically, this equation certainly  does not describe weak transverse propagation. 
In the long wave length regime, the KP equation is well known to describe the propagation of such 
waves when weak transverse perturbation is considered \cite{Dahiya,Kako,Mushtaq}. In an effort to obtain a 2D
extension with weak transverse dependence of modulated ion wave packets propagating in a magnetized plasma, 
 we obtain an asymmetric (2+1) dimensional novel equation(\ref{2dnls})
along with a space-like NLS equation (\ref{spaceNLS}).  Hence, we  observe that our equation has 
resemblance to the KP equation from the point of weak transverse propagation, and to the DS equation
from its modulated structure. 
In case of DS or KP,  the system reduces to NLS or KdV equation
respectively when the transverse coordinate is neglected, but 
the present equation given in (\ref{2dnls}) does not reduce to the standard NLS equation in  such limit, indicating
 its distinctive asymmetric nature.   

The various properties, as well as the solutions of this new equation are explored in 
the foregoing sections of the paper.

\section{Modulation instability}

Instability of a planar wave, appearing due to the interplay between
dispersion and nonlinear effect called modulation instability (MI) \cite{BF},
 which has been in the continuous focus for
many years \cite{McLean,Ramamonjiarisoa}.

  For investigating the contributions to
the frequency due to the linear dispersive and the nonlinear term in
(\ref{2dnls1}), we insert the plane wave solution $ q_0= A_0 \
e^{i(\omega t+k^xx+ k^y y)} ,$ with $A_{0}$
 as the real constant  amplitude, $ \omega$ as frequency and
$(k^x,k^y) $ as the wave vector .  For the plane wave to be
 an exact solution of (\ref{2dnls1}), the frequency should be $ \omega= \omega_L+\omega_{NL},
\ \omega_L=- \frac{C_1}{C_0}k^xk^y, \ \omega_{NL}=\frac{2 C_2}{C_0}A_0^2k^x,$
 where $ \omega_L $
is the frequency due to linear dispersion and $\omega_{NL} $ is its
nonlinear correction, which depends on the amplitude  of the wave as well as
on the x component of the  wave vector.

Now to explore  the onset of MI  in the system
affecting this plane wave solution, we perturb it by a small parameter
function  $\epsilon(x,y,t) $. Note that the perturbation is considered in both the space directions.

 The solution
 \begin{equation}
 q_\epsilon= ( A_0 +\epsilon)
e^{i(\omega t+k^xx+ k^y y)} 
\end{equation}
 neglecting the higher order terms in $\epsilon $  yields from (\ref{2dnls1}) a linear
equation for $\epsilon $ as \begin{equation}  i C_0\epsilon_t+ C_1\epsilon_{xy}+
i C_1(k^y\epsilon_{x}+k^x\epsilon_{y})+
 i C_2
A_0^2(\epsilon_{x}^{*}-\epsilon_{x} )+2 C_2 A_0^2 k^x(\epsilon^*+\epsilon)=0.
\label{epEq}
\end{equation}
 For detecting the
 instability of the perturbation we represent
\begin{equation}
 \epsilon= k_1 e^{i(\omega_{m}t+k^x_{m}x+k^y_{m}y)}+
 k_2 e^{-i(\omega_{m}t+k^x_{m}x+k^y_{m}y)}.
\end{equation}
 Inserting this form of perturbation in equation (\ref{epEq}) and arranging the independent
terms we get a set of two homogeneous equations for the arbitrary
coefficients $k_1,k_2, $ nontrivial solutions of which can exist only when
the determinant  of the matrix vanishes leading to the necessary
relation
 $ { \bar \omega_m}^2=K^2- {\Omega
_c} , $\ 
{where }\
$
  \bar \omega_m= C_0 \omega_m -\omega_{0},$ \ and  
\ $ \omega_{0}= C_2 A_0^2 k_m^x- C_1{k^x}{k_m^y}- C_1{k_m^x}{k^y} ,$
\ $ K= C_1{k_m^x}{k_m^y}-2 C_2 A_0^2 k^x, \  \ \Omega_c=
A_0^4(4 C_2^2 {k^x}^2 -C_2^2{k_m^x}^2),$
which gives finally
\begin{equation}
 C_0 \omega_m=  \omega_0 \pm i \omega_{i} 
 ,\
 \omega_{i} = (\Omega_{c}-K^2 )^\frac{1}{2} = (4 C_2C_1{k_m^x}{k_m^y}k^x A_0^2- C_2^2 A_0^4 {k_m^x}^2
 -C_1^2{k_m^x}^2 {k_m^y}^2)^\frac{1}{2}.
\label{growthrate}
\end{equation}
Therefore,  under the condition $K^2< {\Omega_c} $ which is
 $4 C_2C_1{k_m^x}{k_m^y}k^x A_0^2> C_2^2 A_0^4 {k_m^x}^2
 +C_1^2{k_m^x}^2 {k_m^y}^2$ with
  ${\Omega_c}>0$, i.e when $|{k_m^x}|<2| {k^x}|$ the modulation frequency $
\omega_m$ can acquire an imaginary part $\omega_{i}, $ initiating an
exponential growth of perturbation with time $t$ and hence onsetting the
MI.  $\omega_{i}$ is the growth rate of the instability
given by (\ref{growthrate}),
a graphical form of which is presented in Fig. 2,  showing its dependence on
the longitudinal and transverse directions through $k_{x}^m$, and $k_{y}^m$
respectively. 
For plotting the growth rate and stability region,  specific values of $C_0, C_1$
and $C_2$ are chosen from Figure (1) as -1.65, 0.46 and -0.49 respectively
when $\alpha = 0.4$.
Both these figures show clearly that the behavior of MI as well as growth rate has a strong directional preference 
and range. 

\begin{figure}[!h]
\centering
{
 \includegraphics[width=7cm, angle=0]{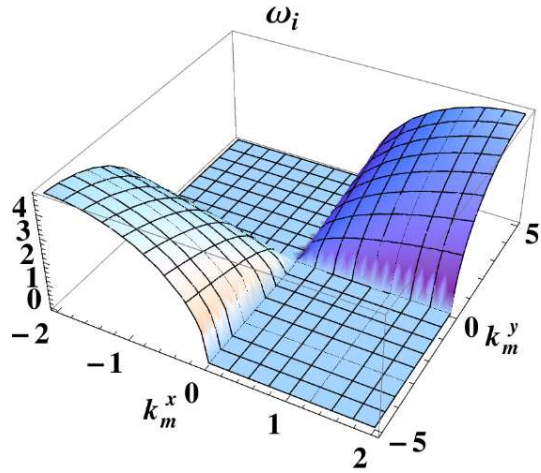}
}
\caption{ The growth rate $\omega_{i}$ of the MI given
by (\ref{growthrate}), arising in our new equation, exhibiting how it
changes (for $A_0=1.0, k^x=-5.0 $) along the longitudinal $(k_m^x)$ and transverse ($k_m^y$)
directions, showing a strong directional preference.} \label{fig:1}
\end{figure}
 The stability plot is drawn in Fig. 3 in the $(k^x_m,k^y_m)$ plane 
with the shaded region
showing the domain of MI.
\begin{figure}[!h]
\centering
{
 \includegraphics[width=7cm, angle=0]{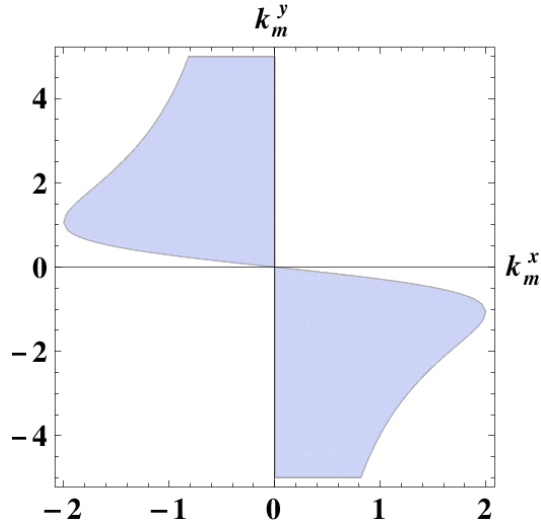}
}
\caption{ Graphical representation of the MI region,
where the instability can occur only within the shaded area (for fixed
values of $A_0=1.0, K^x=-5.0 $).  The
instability region, showing dependence
 on the wave vector $(k_{x}^m ,k_{y}^m
)$, varies asymmetrically along the longitudinal and transverse direction,
as seen clearly from the figure.  } \label{fig:2} \end{figure}

A comparison with these obtained conditions of MI of (\ref{2dnls}) 
with other modulated wave equations may be illuminating. Evolution of a wave satisfying 
one dimensional NLS equation depends on the product of the coefficients of nonlinear and dispersive terms\cite{Moslem},
which are the functions of the physical parameters involved. If the product is positive,
the amplitude modulated envelope is unstable for perturbation wavelengths larger than a critical value, hence
the carrier wave may either collapse or lead to 
formation of bright envelope modulated wave packets.
 On the other hand if the product is negative, the amplitude modulated envelope will be stable against 
external perturbations and may propagate in the form of a dark envelope 
wave packet. For dust acoustic waves and dust ion acoustic waves following NLS equation\cite{Amin},
 amplitude modulation is considered to propagate oblique to 
the direction of the pump carrier wave propagation in \cite{Amin} and
 a wide domain of angle and frequency has been found where
the modulation is unstable. In case of the propagation of
  ion acoustic envelope solitons in cylindrical and spherical geometries following
modified NLS equation in multicomponent plasma with positrons\cite{Sabry}, a modulation instability period 
has been found, which does not exist in one dimension. It is found
that spherical waves are more structurally
stable to perturbations than the cylindrical waves and
increasing positron concentration leads to a decrease in critical wave number until the concentration
reaches a critical value and then further increase of concentration leads the critical wave number to 
increase\cite{Sabry}.

These are the discussions in (1+1) dimension following one space directions. The situation
becomes more complicated and intricate  if the transverse 
perturbations are included. For example, the propagation
of dust ion acoustic waves with combined effect of bounded cylindrical geometry and
transverse perturbations\cite{Xue} leads to cylindrical DS equation. MI features of this system
is widely different from the one dimensional case involving extended parameter domain
and time dependence in instability condition.

Our system ({\ref{2dnls1}}) is also (2+1) dimensional involving asymmetric dependence on the transverse 
coordinate. From the condition (\ref{growthrate}) we see that the growth rate and instability condition
is more complicated involving all longitudinal and transverse components
 $(k^x_m,k^y_m)$ with a strong
directional preference and range $|{k_m^x}|<2| {k^x}|$, compared to the one dimensional case.
The conditions also involve the coefficients $C_1, C_2$ of equation (\ref{2dnls1})
which are the functions of the system parameters
$k_x, k_y$ and $\alpha$. The graphical variation of $C_1, C_2$ with magnetic field $\alpha$ 
for given set of $k_x,k_y$ is shown in Figure-1
from where we can see that numerical value of both
$C_1, C_2$ increases as we increase the magnetic field. Since the product $C_1 C_2$ is negative, modulation instability
can set in when the product $k^x_m k^y_m k^x$ is negative,
and the system remains modulationally stable for positive values of $k^x_m k^y_m k^x$ .
These are the interesting features of our exact model having distinct structure than the (1+1)
dimensional NLS or (2+1) dimensional DS equation.

\section{Connection with Kadomtsev- Petviashvili (KP) equation }
 It is interesting to note that the new integrable equation (\ref{2dnls}) together with the space NLS equation 
 derived in (\ref{spaceNLS}), has a deep connection with another wellknown (2+1) dimensional evolution equation.
The equation (\ref{2dnls}) along with (\ref{spaceNLS}) is a complex equation denoting ion acoustic wave
with modulation. But if we are concerned with the modulus of the wave, then our system
also provides another well known, integrable equation.

Equation (\ref{2dnls}) is our new evolution equation after scaling, while the space NLS equation (\ref{spaceNLS})
at the same scaling becomes
\begin{equation}
 i u_y = B u_{xx} + A|u^2|u,
\label{sspaceNLS}
\end{equation}
where $A, B$ are two real constants dependent on $C_1, C_2$ as
$B= -\frac{B_1^{(3)} C_2}{A_1^{(3)} C_1}$, $A = -\frac{C_1^{(3)} C_1}{A_1^{(3)} C_2}$. 
Now, if (\ref{2dnls}) is multiplied by $u$ and its complex conjugate equation by $u^*$ and subtracted
from one another, then taking derivative w.r.t  $x$ we get another equation
containing quadratic power in $u$. Now using (\ref{sspaceNLS}) and its complex conjugate equation, that equation
can be simplified to yield
\begin{equation}
 [4B \phi_{t}-6AB \phi \phi_{x}-B^2 \phi_{xxx}]_{x} + 3\phi_{yy} = 0,
\label{KP}
\end{equation}
where $\phi = u u^*$.

Equation (\ref{KP}) is nothing but the well known KP equation 
where $\phi$ being the square of modulus of the wave $u$, the dependent variable of equations (\ref{2dnls}) and
(\ref{sspaceNLS}).
It means that the square of the absolute value of the wave without modulation satisfies another 
real equation which
is also integrable and weak in the transverse perturbation.
Note that, in the derivation of (\ref{2dnls}) we have implied weak scaling on the transverse coordinate,
similar as the scaling involved in the derivation of KP equation. Hence our equation (\ref{2dnls})
though being a modulated equation in two dimensions, has a different structure than the DS equation
which is symmetric in both the spatial variables and has a connection with another two dimensional
long wave equation which is asymmetric in the transverse direction.  

For ion acoustic wave, KP equation has been derived and its stability properties
under transverse perturbations have been discussed in many plasma systems like
quantum electron ion plasma\cite{Mushtaq} or in electron-positron-ion plasma with high energy tail 
electron and positron distribution\cite{Shahmansouri} and in other fields and using Sagdev potential approach
conditions of existance of stable solitary waves have been obtained. Unlike KdV equation
where the form of soliton solution does not depend on the sign of the dispersion term, the form
of soliton solution of KP equation is directly determined by the dispersion sign. Again the stability of
the soliton depends on the coefficients of the KP equation which in this case also depends on
the coefficients of (\ref{2dnls1}) and (\ref{spaceNLS}).
 Depending on the signs
of $A,B$ equation (\ref{KP}) can be transformed into KP-I or KP-II equations 
admitting different types of solutions .
The square root of various solutions of (\ref{KP})
can be used for constructing the solution of (\ref{2dnls}) together with a phase factor connected with
(\ref{sspaceNLS}), which we plan to explore in the future. 
The connection of our equation (\ref{2dnls}) with KP equation stresses the two dimensional and asymmetric
nature of the equation.

\section{Soliton  solutions} 

As a direct feature of integrable system, our equation, 
  (\ref{2dnls})  must admit higher soliton solutions.  In this section we will elaborately discuss its multisoliton solutions, which can be derived by many methods e.g, inverse scattering transform, Hirota method and various dressing methods. The IST method is more powerful (it can handle general, initial conditions) and at the same time more complicated.  But, if one just wants to find soliton
  solutions, Hirota's method is fastest in producing results \cite{Hirota,Lakshmanan}.  Hence,
 following the method, we use the bilinearizing transformation given by
\begin{equation}
 u(x,y,t)= \frac{G(x,y,t)}{F(x,y,t)}, \label{GF} \end{equation} where
$G(x,y,t)$ and
$F(x,y,t)$ are complex and real functions respectively. 
 Using (\ref{2dnls}) one derives  a  pair of bilinear
equations:
 \begin{equation}
 i(FG_{t}-GF_{t}) + (FG_{xy}+ GF_{xy}-G_{x}F_{y}-G_{y}F_{x})=0,  2i(GG_{x}^*-G^*G_{x}) + 2(F_{x}F_{y}-FF_{xy})=0.
\label{Hirota2}\end{equation}
Multisoliton solutions are obtained by finite perturbation expansions as 
\begin{equation}
  F = 1 + \epsilon^2 F_{2} + \epsilon^4 F_{4} + \cdots,
G =  \epsilon G_{1} + \epsilon^3 G_{3} +  \cdots,  \label{G123} 
\end{equation}

where   $\epsilon$  is formal expansion parameter need not  to be small.
  Collecting like powers of $\epsilon$, we obtain the following series of equations:
\begin{equation}
{\rm O}(\epsilon):\ \  iG_{1t} + G_{1xy}=0
\label{G1}\end{equation}
\begin{equation}
{\rm O} (\epsilon^2): \ \  2F_{2xy}=2i[G_{1}G_{1x}^* - G_{1}^*G_{1x}]
\label{F2}\end{equation}
\begin{equation}
{\rm O} (\epsilon^3): \ \ iG_{3t}+G_{3xy}=
i[G_{1}F_{2t}-F_{2}G_{1t}]-[F_{2}G_{1xy}+G_{1}F_{2xy}-G_{1x}F_{2y}-G_{1y}F_{2x}]=0
\label{G3}\end{equation} \begin{equation}
{\rm O} (\epsilon^4): \ \  2F_{4xy}=2i[G_{3}G_{1x}^*
+G_{1}G_{3x}^*-G_{3}^*G_{1x}-G_{1}^*G_{3x}]+2[F_{2x}F_{2y}-F_{2}F_{2xy}]
\label{F4}\end{equation} and similarly  higher order equations.

  \subsection { 1-soliton}
To construct    1-soliton solution for 
(\ref{2dnls}) we assume the ansatz
\begin{equation}
 G_{1}=e^{\eta_{1}}, \ 
 \eta_{1}=k_{1}x + p_{1}y-w_{1}t+\eta_{1}^{0}\label{G11}
\end{equation}
where $k_{1}, p_{1},w_{1},\eta_{1}^{0}$ are complex constants.
 From equation (\ref{G1}) therefore one obtains
 the associated dispersion relation
$\  w_{1}=-ik_{1}p_{1}, \ $
using which  the equation  (\ref{F2}) is solved easily  to yield 
\begin{equation}
 F_{2}=i(k_{1}^*-k_{1})
\frac{e^{(\eta_{1}+\eta_{1}^*)}}{(p_{1}+p_{1}^*)(k_{1}+k_{1}^*)}. \label{F22}
\end{equation} 
We can verify using  (\ref{G11}) and  (\ref{F22}), that
 all  higher order terms in $\epsilon$ 
beyond $G_{1}$ and $F_{2}$  trivially vanish.
  Absorbing $\epsilon$ in   arbitrary
 constant  $\eta_{1}^{0}$, 
we construct from  
  (\ref{GF}) using  (\ref{G11}) and  (\ref{F22})
the 1 soliton solution in the form 
\begin{equation}
 u(x,y,t)=\frac{G_{1}}{1+F_{2}}=\frac {e^{\eta_{1}}}
 {1+ \alpha 
e^{(\eta_{1}+\eta_{1}^*)}}
\label{1sol0} \end{equation} 
where $\alpha$ depends on the parameter $k_{1}, p_{1} .$ 
One can identify the interesting 2d nature of our equation (\ref{2dnls}) by making $p_1=0$, then from eq. (\ref{F22}) we can see that $F_2$ will diverge and no soliton solution can be found.
If additionally we use the dispersion relation of the constraint  equation 
\begin{equation}
 i u_y + u_{xx} + 2|u|^2 u = 0
\label{const}
\end{equation}
 which comes from (\ref{spaceNLS}),
  as  $ p_{1} = -i k_{1}^2$,  the 
soliton solution (\ref{1sol0}) simplifies to
 yield the conventional form 
\begin{equation}
 q(x,y,t)
={\rm sech}\xi \ e^{i\theta}, \ \mbox{with} \ \xi=\eta(x+v_y y+v t), \
\theta=(k_xx+k_yy+\omega t).  \label{1sol2}
\end{equation}
       
A frozen picture of the modulus of our travelling  soliton solution (\ref{1sol2}) at
time $t=0 $  is shown in Fig. 4.

\begin{figure}
\includegraphics[height=5cm,width=7cm]{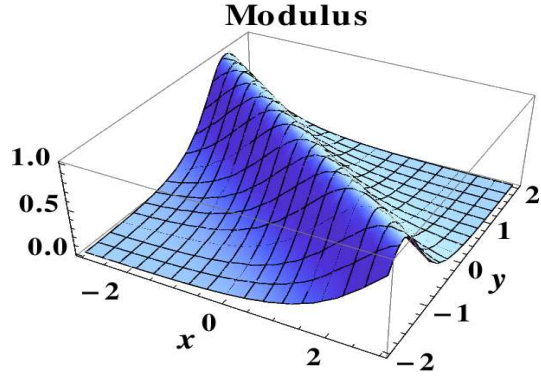}
\caption{Modulus of 1 soliton with $k_{1r}=1,k_{1i}=1,\eta_{1r}^{0}=1,
\eta_{1i}^{0}=-1 $ and at t=0} 
\end{figure}
\subsection{2-Soliton}

For  obtaining  2-soliton solution we start with  the standard procedure
assuming  
\begin{equation}
 G_{1}=e^{\eta_{1}}+ e^{\eta_{2}},
 \eta_{1}=k_{1}x + p_{1}y-w_{1}t+\eta_{1}^{0}
, \  \eta_{2}=k_{2}x + p_{2}y-w_{2}t+\eta_{2}^{0} 
,
\label{G12}\end{equation}
 where the parameters   involved are complex numbers. 
 Applying similar   dispersion relations as earlier 
we get 
$ w_{1}=-ik_{1}p_{1}, \ 
 w_{2}=-ik_{2}p_{2} $ and obtain from (\ref{F2})
 \begin{equation}
 {F_{2}=[e^{(\eta_{1}+\eta_{1}^*+R_{1})}+
 e^{(\eta_{2}+\eta_{2}^*+R_{2})}}
+e^{(\eta_{2}+\eta_{1}^*+\delta_{0})}+
{e^{(\eta_{1}+\eta_{2}^*+\delta_{0}^*)}]},
\label{2F2}\end{equation}
where all the constant parameters can be worked out explicitly
 (see Appendix 
).
Similarly
 equation (\ref{G3}) at  higher order expansion   gives 
\begin{equation} {G_{3} = e^{(\eta_{1}+\eta_{1}^*+\eta_{2}+ \delta_{1})} 
+ e^{(\eta_{1}+\eta_{2}^*+\eta_{2}+ \delta_{2})}},
\label{2G3}\end{equation}
 where the relevant 
 parameter details 
are given in  Appendix .  
 Using  further equation (\ref{F4}) one obtains
\begin{equation}
  F_{4} = e^{(\eta_{1}+\eta_{1}^*+\eta_{2}+\eta_{2}^*+R_{3})},
\end{equation}
 with the relevant parameters
 presented in  Appendix. 
For simplifying the expressions,as mentioned earlier, 
we can use the constraint equation (\ref{const}),
imposing the relations between
 $k_{1}$ , $p_{1}$ and $k_{2}$ , $p_{2}$ as $p_{1}=-ik_{1}^2$ ,
 $p_{2}=-ik_{2}^2$ (see Appendix).  Here we find again, that the higher
 order terms in $\epsilon $ beyond $G_{3}$ and
$F_{4}$ trivially vanish, leaving the exact 2- soliton solution in the form
\begin{equation}
 u(x,y,t)=\frac{G_{1}+G_{3} }{1+F_{2}+ F_{4}}\label{2sol} \end{equation} 
 Here also we can see that for no transverse dependence i.e. $p_1=p_2=0$, all the quantities determined $F_2,G_3,F_4$
 diverges and two soliton solution cannot be found  which indicates the strict 2d nature of the equation.
 A
graphical plot of the  modulus of this solution   in $(2+1)
$-dimensions, frozen at time $t=2$, is shown in Fig.  5, where 
the 2-soliton as two interacting 1-solitons 
 is clearly seen on a 2D $(x,y) $-plane. Following the same procedure the higher 
soliton solutions of our equation can be evaluated.

These one and two soliton solutions of (\ref{2dnls}) have similarities with the soliton solutions
of NLS equation with an additional transverse dependence. Since a purely 1D model cannot account for the
observed features of many physical situations, specially in auroral region
with higher polar altitudes\cite{Franz} corresponding 2D model is necessary. There lies the importance of the 
soliton solutions of our equation (\ref{2dnls}), which could be applied to many research areas of this field.

\begin{figure} \includegraphics[height=5cm,width=7cm]{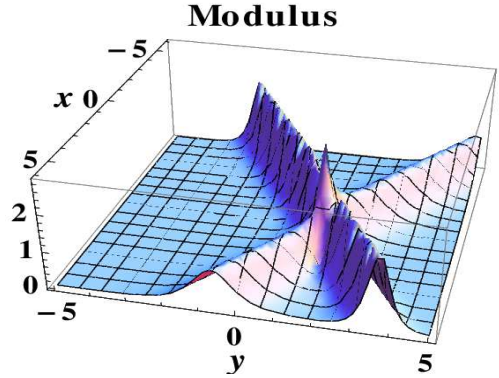}
\caption{Modulus of 2 soliton with
$k_{1r}=1,k_{1i}=1,k_{2r}=2,k_{2i}=-1,\eta_{1r}^{0}=1, \eta_{1i}^{0}=1,
\eta_{2r}^{0}=1, \eta_{2i}^{0}=1 $ and at t=2} \end{figure}

\section{ Exact static 2D lump  solution:}
Localized wave structure 
in (2+1) dimensional systems are very important in terms theoretical
and experimental aspects of plasma. Localized rational structure
following KP-I equation have been found by Janaki et.al in the  propagation of oblique magnetosonic
wave in warm collisional plasma system\cite{Janaki}, where the wave profile decays algebraically in both directions.
Whereas, in case of modulated wave equation like Davey Stewartson system, an exponentially localized solution
called Dromion, which moves with time has been found
in magnetized electron acoustic wave system \cite{Ghosh}. 
 Our equation (\ref{2dnls}) which has some similarity with both DS and KP equations, also 
possesses an exact  localized wave solution, decaying rationally in both spatial directions.
   
The static 2D rational lump solution is given by
\begin{equation}
 u_{static}(x,y)= exp(4iy)(\frac{1-4iy}{c+\alpha_1 x^2+4 y^2}-1)
\label{staticlump}
\end{equation} where c , $\alpha_1$ are 2 free parameters.  From this we can see that the wave attains the
maximum amplitude
 \begin{equation}
 A_{max}=\frac{1-c}{c} \end{equation} at the centre x=0, y=0 which can be
controlled by c.  At large distances($|x|$ $\rightarrow \infty $,$|y|$
$\rightarrow \infty $)
 the amplitude goes to unity. The steepness of this static wave solution as
 observed from the front is $ \frac{\partial}{\partial y}{u_{static}}$ is
 related to another free parameter $\alpha_1$.The amplitude of the wave falls
 to its minimum at x=0, $y= \pm y_{0}$, with
$y_{0}=\sqrt{\frac{1-c}{\alpha_1}}$.
Hence the density gets localized at the centre x=0,y=0 and the concentration
can be controlled by the free parameter $c$. This is an interesting feature
 because in actual physical situation the ion
density can change which need to be controlled by the free parameters.
This is absent in the (1+1) dimensional NLS system, where Peregrine breather which is used
to describe density localization in the x-t plane, have no free parameters and hence can
achieve a fixed maximum value (3 times than the background).
Whereas our solution, having two free parameters $c$ and $\alpha_1$
 can achieve any amplitude and steepness, relevant to the actual physical condition.

Thus unlike the exponentially decaying dromion solution of DS1 equation,
our system (\ref{2dnls}) provides a rational solution in both space directions having similar structure to the
rational solution of KP-I equation. There lies another  connection of the equation (\ref{2dnls})
with KP which has similar scaling and weaker dependence on transverse directions
as (\ref{2dnls}).
 Unfortunately, the time evolution of the rational solution (\ref{staticlump})
 could not be found which can be explored in future.

\begin{figure}
\includegraphics[height=7cm,width=7cm]{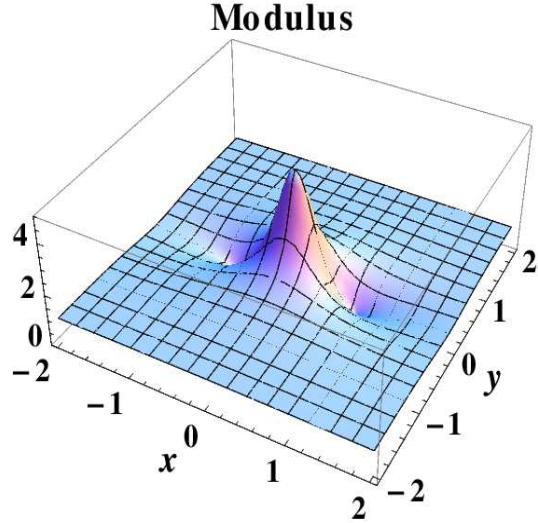}
\caption{Absolute value of the solution (\ref{staticlump}) with c=1/6, $\alpha=6/5 $ }
\end{figure}

\section{Other integrable properties: Lax Pair, Conserved quantities}
Since our new equation (\ref{2dnls}) is a completely integrable system, it possesses 
the associated integrable properties. One of
such properties i.e, existence of higher soliton solutions, have been discussed 
using Hirota bilinearization method in the previous section.
We briefly state  here the other properties such that existence
of Lax pairs, infinite number of conserved quantities etc \cite{mypaper}.
 
 For equation (\ref{2dnls}), one can find
 the associated linear  system
 $\Phi_y=U(\lambda)\Phi, \Phi_t=V(\lambda)\Phi, $  with a Lax pair given by
\begin{equation} U(\lambda )\equiv V_2(\lambda)=2\lambda V_1(\lambda)+V_2^{(0)},\
V(\lambda)\equiv V_3(\lambda)=2\lambda V_2(\lambda)+V_3^{(0)}
\label{V23}\end{equation}
where 
\begin{eqnarray}
V_1(\lambda)=i(\lambda \sigma^3+U^{(0)}), \
V_2^{(0)}=\sigma^3(U^{(0)}_x-i{U^{(0)}}^2)\nonumber\\
V_3^{(0)}=
-\sigma^3U^{(0)}_y -[U^{(0)},U^{(0)}_x], \
U^{(0)}= q \sigma^+ + q^* \sigma^-
, \label{Lax}
\end{eqnarray}
with $\sigma^a, a = \pm,3, $ Pauli matrices.
 The flatness condition: $U_t-V_y+[U,V]=0$, of the given Lax Pair gives
 our equation
(\ref{2dnls}) together with the space NLS equation like (\ref{sspaceNLS}) for particular values of $A$ and $B$. The derivation
done in sec II from the two fluid equations for a magnetized plasma has shown to lead to  the space NLS as well as the 2d evolution equation(\ref{2dnls}) at different orders of perturbation,
 which is consistent with the Lax pair formalism.

Systems with infinite degrees of freedom like 
(\ref{2dnls}), when integrable, 
should have infinite set of independent  conserved quantities, which can be derived explicitly as 
\begin{eqnarray} C_1= i\int dy (q^*q_x-q^*_xq),   \  C_2=\int dy
  (i\frac 1 2 (q^*_{y}q -q^*q_{y})+ q^*_{x}q_{x} + | q|^4
) , \ C_3=\int dy (q^*_{y}q_x+q^*_{x}q_y) ,\nonumber \\  
   C_4= \int dy \left(
iq^*_{xy}q_{x}+q^*_{y}q_y-
i    |q|^2(q^*q_{y}- q^*_yq)
 -2 |q|^2 \ q^*_xq_{x}+( {q^*}^2q^2_{x}+
{q_x^*}^2q^2
 )\right),\label{2dCn}\end{eqnarray}
and so on.
Note the involvement of both the space-variables $x,y $ in this series of 
independent conserved quantities, which also gives another 
 argument in favor of the integrability of the 2D nonlinear 
 equation (\ref{2dnls}). Nevertheless, the one dimensional integral of these conserved quantities signifies
also about,a quasi-2D nature of this system.

\section{Conclusive remarks}
In this work a completely integrable, (2+1) dimensional, modulated , nonlinear evolution equation has been derived
in the ion acoustic wave of magnetized collisionless plasma system.
It has been obtained at a higher 
perturbation order compared to the NLS equation, hence expected to address weaker effects.
The two spatial directions are scaled asymmetrically allowing weak transverse perturbation.
Thus the equation derived (\ref{2dnls}), has connection with the DS equation in terms 
of its modulated structure, with KP equation in terms of its weak transverse dependence.
Condition of modulation instability and nonlinear frequency shift
clearly shows the spatial asymmetry in growth rate, which is a distinct feature compared to the
 one dimensional case.  Dependence of the onset condition of MI 
with magnetic field and other system parameters are discussed.
It is shown that the square of modulus of the dependent variable of the equation satisfies
KP equation along with the space NLS like constraint(\ref{spaceNLS}) equation.
Its Lax Pair structure is discussed and infinite conserved
quantities are derived briefly. Using Hirota bilinearization scheme its higher soliton solutions are calculated
which has similar structure with the 1D soliton with an additional
dependence on transverse direction.
The exact algebraic lump solution of (\ref{2dnls}), carrying 2 free parameters is also  given
and density localization of the system is discussed.
Applications
of this important and novel equation in various aspects of plasmas and identification of its different features may 
pave a new direction of research in this field.

\section {Appendix: }

\noindent {\bf i) } 
{\bf Coefficients appearing in the derivation of the two dimensional integrable evolution equation
(\ref{2dnls}),
given in section-II:}

\vspace{0.5 cm}

$ K_{1} = \frac{1}{1 + |k|^2}$, $|k|^2 = (k_{x}^2 + k_{y}^2) $,$\kappa_{1} = \frac{1}{1+4|k|^2}$,
$A_{1}^{(1)} = -\frac{(i K_{1} k_{y}\alpha + K_{1} k_{x} \omega)}{(\alpha^2-\omega^2)}$,
$B_{1}^{(1)} = \frac{(i K_{1} k_{x}\alpha - K_{1} k_{y} \omega)}{(\alpha^2-\omega^2)}$
\vspace{0.5 cm}

$A_{0}^{(2)} = \frac{- 2 \omega k_{x}}{|k|^2} $,
$B_{0}^{(2)} = \frac{- 2 \omega k_{y}}{|k|^2} $,
\vspace{0.5 cm}

$A_{1}^{(2)} = -\frac{(i K_{1} k_{y} \alpha + K_{1} k_{x} \omega)}{(\alpha^2 - \omega^2)}$,
$C_{1}^{(2)} = -\frac{(-i K_{1} k_{x} \alpha^3 + K_{1} k_{y} \alpha^2 \omega + 
  i K_{1} k_{x} \alpha \omega^2 - 
  K_{1} k_{y} \omega^3)}{(-\alpha^2 + \omega^2)^2}$,

\vspace{0.5 cm}
$D_{1}^{(2)} = -\frac{(-i K_{1} k_{y} M_{x} \alpha^2 - K_{1} \alpha^3 + 2 K_{1}^2 k_{x}^2 \alpha^3 - 
  2 K_{1} k_{x} M_{x} \alpha \omega + 2 i K_{1}^2 k_{x} k_{y} \alpha^2 \omega - 
  i K_{1} k_{y} M_{x} \omega^2 + K_{1} \alpha \omega^2 - 
  2 K_{1}^2 k_{x}^2 \alpha \omega^2 - 
  2 i K_{1}^2 k_{x} k_{y} \omega^3)}{(-\alpha^2 + \omega^2)^2}$,

\vspace{0.5 cm}
$B_{1}^{(2)} = -\frac{\{-2 K_{1}^2 k_{x} k_{y} \alpha - i K_{1} \omega + 2 i K_{1}^2 k_{x}^2 \omega - (
  i K_{1} k_{x} M_{x} \alpha^2)/(\alpha^2 - \omega^2) + (
  2 K_{1} k_{y} M_{x} \alpha \omega)/(\alpha^2 - \omega^2) - (
  i K_{1} k_{x} M_{x} \omega^2)/(\alpha^2 - \omega^2)\}}{(\alpha^2 - \omega^2)}$,
\vspace{1 cm}

$A_{2}^{(2)} = \{-2 K_{1}^3 k_{x} |k|^4 \kappa_{1} + 
   K_{1} |k|^2(5 k_{x} \alpha^2 + 
      3 i k_{y} \alpha \omega  + 
      4 k_{x}^3 \kappa_{1} + \nonumber \\
 4 k_{x} k_{y}^2 \kappa_{1}) + \alpha^2 (3 k_{x} \alpha^2 + 
      3 i k_{y} \alpha \omega + 
      4 k_{x} |k|^2 \kappa_{1}) + 
   K_{1}^2 |k|^2 (2 k_{x} |k|^2 - 
      2 k_{x} \alpha^2 \kappa_{1} - 
      i k_{y} \alpha \omega \kappa_{1})\}/\{3 \alpha^2 \omega |k|^2 + 4 |k|^4
      \omega (K_1- \kappa_1)\}$,

\vspace{0.5 cm}

$B_{2}^{(2)} = \{-2 K_{1}^3 k_{y} |k|^2 \kappa_{1} + 
   K_{1} |k|^2 (5 k_{y} \alpha^2 - 
      3 i k_{x} \alpha \omega + \nonumber \\
 4 k_{x}^2 k_{y} \kappa_{1} + 
      4 k_{y}^3 \kappa_{1}) + \alpha^2 (3 k_{y} \alpha^2 - 
      3 i k_{x} \alpha \omega + 
      4 k_{y} |k|^2 \kappa_{1}) + 
   K_{1}^2 |k|^2 (2 k_{y} |k|^2 - 
      2 k_{y} \alpha^2 \kappa_{1} + 
      i k_{x} \alpha \omega )\}/\{3 \alpha^2 \omega |k|^2 + 4 |k|^4
      \omega (K_1- \kappa_1)\} $,

 \vspace{0.5 cm}
$C_{2}^{(2)} =2 \frac{\{3 \alpha^2 + K_{1} |k|^2 (3 - K_{1} \kappa_{1})\}}{\{
3 \alpha^2 + 4 |k|^2 (K_{1} - \kappa_{1})\}}$,

\vspace{0.5 cm}
$D_{2}^{(2)} = \frac{\{-4 K_{1} (-3 + K_{1}^2) |k|^2 - 
   3 (-4 + K_{1}^2) \alpha^2\} \kappa_{1}}{\{
6 \alpha^2 + 8 |k|^2 (K_{1} - \kappa_{1})\}}$,

\vspace{0.5 cm}

$A_{1}^{(3)} = 2  K_1 k_y \omega (|k|^2 K_1 - 1) $,

\vspace{0.5 cm}

$B_{1}^{(3)} = \frac{-K_{1} (K_1 |k|^2 -1)\{K_1 (-k_y^2 + 3K_1 k_x^2 |k|^2)+ (4K_1 k_x^2 -1)\alpha^2\}}{\omega}$,
\vspace{0.5 cm}

$C_{1}^{(3)} = |k|^2 \omega [K_1 \{4K_1 (K_1^3 -3)|k|^2 + 3(K_1^3-4)\alpha^2\}+\{4 K_1(12 + K_1^2(K_1 -3)(K_1 +2)\}|k|^2
+ 3(k_1^2-4)^2 \alpha^2\}\kappa_1]/
\{6 \alpha^2 + 8|k|^2 (K_1 - \kappa_1)\}$,
\vspace{1 cm}

$A_0^{(3)} = -\frac{1}{\alpha}[A_{1}^{(1)*} B_{1}^{(1)}-i B_{1}^{(2)} B_{1}^{(1)*}k_x + i A_{1}^{(1)*} D_{1}^{(2)}k_x
- B_{0}^{(2)} M_x]$

\vspace{0.5 cm}
$B_0^{(3)} = -\frac{1}{\alpha}[A_{1}^{(1)} B_{1}^{(1)*}-i B_{1}^{(1)} B_{1}^{(2)*}k_x - i A_{1}^{(1)} D_{1}^{(2)*}k_x
- B_{0}^{(2)} M_x]$

\vspace{0.5 cm}
$C_0^{(3)} = \frac{i}{\alpha}[A_{1}^{(2)} B_{1}^{(1)*}- A_{1}^{(1)*} C_{1}^{(2)}]k_x $

\vspace{0.5 cm}
$D_0^{(3)} = -\frac{i}{\alpha}[A_{1}^{(2)*} B_{1}^{(1)}- A_{1}^{(1)} C_{1}^{(2)*}]k_x $

\vspace{0.5 cm}
$E_0^{(3)} = \frac{1}{\alpha}[A_{1}^{(1)} A_{1}^{(1)*}- K_1^2+i B_{1}^{(2)} B_{1}^{(1)*}k_y - i A_{1}^{(1)*} D_{1}^{(2)}k_y
- A_{0}^{(2)} M_x]$

\vspace{0.5 cm}
$F_0^{(3)} = \frac{1}{\alpha}[A_{1}^{(1)} A_{1}^{(1)*}- K_1^2-i B_{1}^{(1)} B_{1}^{(2)*}k_y + i A_{1}^{(1)} D_{1}^{(2)*}k_y
- A_{0}^{(2)} M_x]$

\vspace{0.5 cm}
$G_0^{(3)} = \frac{i}{\alpha}[A_{1}^{(2)} B_{1}^{(1)*}- A_{1}^{(1)*} C_{1}^{(2)}]k_y $

\vspace{0.5 cm}
$H_0^{(3)} = -\frac{i}{\alpha}[A_{1}^{(2)*} B_{1}^{(1)}- A_{1}^{(1)} C_{1}^{(2)*}]k_y $

\vspace{0.5 cm}
$L_0^{(3)} = \frac{1}{M_x \alpha}[A_{1}^{(1)} \alpha +A_{1}^{(2)*}\alpha + i A_{1}^{(1)}C_{1}^{(2)*}k_x -
i A_{1}^{(2)*}B_{1}^{(1)}k_x ]$

\vspace{0.5 cm}
$K_0^{(3)} =\frac{1}{M_x \alpha}[A_{1}^{(2)} \alpha +A_{1}^{(1)*}\alpha + i A_{1}^{(2)}B_{1}^{(1)*}k_x -
i A_{1}^{(1)*}C_{1}^{(2)}k_x ] $

\vspace{0.5 cm}
$J_0^{(3)} = \frac{1}{M_x \alpha}[-A_{1}^{(1)} B_{1}^{(1)*}-i B_{1}^{(1)} B_{1}^{(2)*}k_x + i A_{1}^{(1)} D_{1}^{(2)*}k_x
+ B_{0}^{(2)} M_x + B_{1}^{(2)*}\alpha] $

\vspace{0.5 cm}
$I_0^{(3)} = \frac{1}{M_x \alpha}[-B_{1}^{(1)} A_{1}^{(1)*}+i B_{1}^{(1)*} B_{1}^{(2)}k_x - i A_{1}^{(1)*} D_{1}^{(2)}k_x
+ B_{0}^{(2)} M_x + B_{1}^{(2)}\alpha] $

\vspace{0.5 cm}
$Q_0^{(3)} = -\frac{1}{M_x \alpha}[iA_{1}^{(2)*} B_{1}^{(1)}k_x -i A_{1}^{(1)} C_{1}^{(2)*}k_x -  A_{1}^{(1)} \alpha
-A_{1}^{(2)*}\alpha + K_1^2 M_x \alpha]$

\vspace{0.5 cm}
$P_0^{(3)} = \frac{1}{M_x \alpha}[-A_{1}^{(1)} B_{1}^{(1)*}-i B_{1}^{(1)} B_{1}^{(2)*}k_x + i A_{1}^{(1)} D_{1}^{(2)*}k_x
+ B_{0}^{(2)} M_x + B_{1}^{(2)*}\alpha + 2iK_1^3 k_x M_x \alpha]$

\vspace{0.5 cm}
$M_0^{(3)} = \frac{1}{M_x \alpha}[-A_{1}^{(1)*} B_{1}^{(1)}-i B_{1}^{(2)} B_{1}^{(1)*}k_x - i A_{1}^{(1)*} D_{1}^{(2)}k_x
+ B_{0}^{(2)} M_x + B_{1}^{(2)}\alpha - 2iK_1^3 k_x M_x \alpha] $

\vspace{0.5 cm}
$N_0^{(3)} =- \frac{1}{M_x \alpha}[-iA_{1}^{(2)} B_{1}^{(1)*}k_x+i A_{1}^{(1)*} C_{1}^{(2)}k_x -  A_{1}^{(2)} \alpha
-A_{1}^{(1)*}\alpha + K_1^2 M_x \alpha] $

\vspace{0.5 cm}

 \noindent {\bf ii)} 
{\bf Coefficients appearing in the Hirota bilinearization procedure in solving equation
 (\ref{2dnls}), given in section-V:} 

\vspace{0.5 cm}
$$e^{R_{1}}=i\frac{(k_{1}^*-k_{1})}{(p_{1}+p_{1}^*)(k_{1}+k_{1}^*)},e^{R_{2}}=i\frac{(k_{2}^*-k_{2})}{(p_{2}+p_{2}^*)(k_{2}+k_{2}^*)},
$$ 
$$ e^{\delta_{0}}=i\frac{(k_{1}^*-k_{2})}{(p_{2}+p_{1}^*)(k_{2}+k_{1}^*)},
e^{\delta_{0}^*}=i\frac{(k_{2}^*-k_{1})}
 {(p_{1}+p_{2}^*)(k_{1}+k_{2}^*)}$$

\begin{eqnarray}
 e^{\delta_{1}}&=&
    \frac{i}{[(k_{2}+k_{1}^*)(p_{1}+p_{1}^*)+(k_{1}+k_{1}^*)(p_{2}+p_{1}^*)]}[\frac{(k_{1}^*-k_{1})
    (p_{2}-p_{1})}{(p_{1}+p_{1}^*)}+\nonumber \\
&+&\frac{(k_{1}^*-k_{1})(k_{2}-k_{1})}{(k_{1}+k_{1}^*)}+
\frac{(k_{1}^*-k_{2})(p_{1}-p_{2})}{(p_{2}+p_{1}^*)}+
\frac{(k_{1}^*-k_{2})(k_{1}-k_{2})}{(k_{2}+k_{1}^*)}] \end{eqnarray}
\begin{eqnarray}
 e^{\delta_{2}}&=& \frac{i}{[(k_{1}+k_{2}^*)(p_{2}+p_{2}^*)+(k_{2}+k_{2}^*)(p_{1}+p_{2}^*)]}[\frac{(k_{2}^*-k_{2})
    (p_{1}-p_{2})}{(p_{2}+p_{2}^*)}+\nonumber \\
&+&\frac{(k_{2}^*-k_{2})(k_{1}-k_{2})}{(k_{2}+k_{2}^*)}+
\frac{(k_{2}^*-k_{1})(p_{2}-p_{1})}{(p_{1}+p_{2}^*)}+
\frac{(k_{2}^*-k_{1})(k_{2}-k_{1})}{(k_{1}+k_{2}^*)}]
\end{eqnarray}

\begin{eqnarray} e^{R_{3}}&=&\frac{1}{(k_{1}+k_{1}^*+k_{2}+k_{2}^*)(p_{1}+p_{1}^*+p_{2}+p_{2}^*)}
[\{ie^{\delta_{2}}(k_{1}^*-k_{1}-k_{2}-k_{2}^*)\nonumber
\\
&+&ie^{\delta_{1}}(k_{2}^*-k_{1}-k_{2}-k_{1}^*)+ie^{\delta_{1}^*}(k_{1}^*+k_{2}^*+k_{1}-k_{2})+
ie^{\delta_{2}^*}(k_{1}^*+k_{2}^*+k_{2}-k_{1})\}\nonumber \\
&+&\{(k_{2}+k_{2}^*-k_{1}-k_{1}^*)
(p_{1}+p_{1}^*)+(k_{1}+k_{1}^*-k_{2}-k_{2}^*)
(p_{2}+p_{2}^*)\}e^{R_{1}+R_{2}}\nonumber
\\
&+&\{(k_{2}+k_{1}^*-k_{1}-k_{2}^*)
(p_{1}+p_{2}^*)+(k_{1}+k_{2}^*-k_{2}-k_{1}^*)
(p_{2}+p_{1}^*)\}e^{\delta_{0}+\delta_{0}^*}
]
\end{eqnarray}
  
For simplifying the expressions we can impose  the relations between
 $k_{1}$ , $p_{1}$ and $k_{2}$ , $p_{2}$ as
 $p_{1}=-ik_{1}^2$ , $p_{2}=-ik_{2}^2$, which would yield 
 $$e^{R_{1}}=\frac{1}{(k_{1}+k_{1}^*)^2},e^{R_{2}}=\frac{1}{(k_{2}+k_{2}^*)^2},
e^{\delta_{0}}=\frac{1}{(k_{1}+k_{2}^*)^2},
e^{\delta_{0}^*}=\frac{1}{(k_{2}+k_{1}^*)^2},
 $$ $$
e^{\delta_{1}}= \frac{(k_{1}-k_{2})^2}{(k_{1}+k_{1}^*)^2(k_{2}+k_{1}^*)^2},
e^{\delta_{2}}= \frac{(k_{2}-k_{1})^2}{(k_{2}+k_{2}^*)^2(k_{1}+k_{2}^*)^2},
 $$ $$
 e^{R_{3}}=\frac{|(k_{1}-k_{2})|^4}{(k_{1}+k_{1}^*)^2(k_{2}+k_{2}^*)^2 (|(k_{1}+k_{2}^*|)^4}
$$

\end{document}